\documentclass[aps,twocolumn]{revtex4}
\usepackage{graphicx}
\usepackage{dcolumn}
\begin{document}
\def\be{\begin{equation}}
\def\ee{\end{equation}}
\def\lag{\langle}
\def\rag{\rangle}
\title{Slow regions percolate near glass transition} 
\author{ Y. Y\i lmaz$^{1}$, A. Erzan$^{1,2}$, and \"O. Pekcan$^1$}
\affiliation{$^1$  Department of Physics, Faculty of  Sciences
and
Letters\\
Istanbul Technical University, Maslak 80626, Istanbul, Turkey }
\affiliation{$^2$  G\"ursey Institute, P. O. Box 6, \c
Cengelk\"oy 81220, Istanbul, Turkey}
\date{\today}
\begin{abstract}
A nano-second scale {\it in situ} probe reveals that a bulk linear polymer undergoes a sharp phase transition as a 
function of the degree of conversion, as it nears the glass transition. The scaling behaviour is in the same universality class as percolation. The exponents 
$\gamma$ and $\beta$ are found to be $1.7 \pm .1$ and $0.41\pm 0.01$ in 
agreement with the best percolation results in three dimensions.

\end{abstract}
\maketitle

\section{Introduction}

Polymeric melts, or linear polymers in the bulk, exhibit a rich array of 
different behaviours,  as a function of the 
temperature, specific volume (per chain), chain length (molecular weight), and 
timescale of observation, 
generally without sharp boundaries between the different regimes. 
Linear polymers in the bulk range from solid, glassy or  leathery materials at low 
temperatures to  rubbery, and eventually viscous fluids at higher 
temperatures or polymer to solvent volume ratios~\cite{deGennes1,Doi}. 
The viscosity of bulk linear polymers typically increases with increasing chain 
length and/or degree of conversion as polymerization proceeds, and diverges near 
the glass transition~\cite{Goetze}. The glass transition temperature is a function of the average chain length, for small to moderate length chains, but ceases to depend on the molecular weight for relatively long chains.
The rheological or viscoelastic properties can thus be studied experimentally either as a function of the temperature, or as a function of the reaction time during the conversion process, as we have done here.  

In this paper we would like to report that the extremely short-time
response of bulk linear polymers probed by a fluorescent aromatic molecule
displays a sharp transition as a function of the reaction time, which is a
(nonlinear) measure of the degree of conversion. Immediately after the
transition point, the melt is found to be glassy or rubbery for
temperatures respectively below and above the glass transition temperature
quoted for commercially available bulk PMMA~\cite{kemco}, namely
106$^\circ$C.

We find that
this transition is characterised by
percolation critical exponents.
In an unpublished paper~\cite{comment1},
we interpreted, erroneously, the percolation exponents found 
here as signalling 
the onset of the entanglement percolation transition. We now believe, for 
reasons to be discussed in this paper, that the fluorescence signal 
cannot, in fact, detect the formation of entanglement clusters.

In order to understand the physical nature of the processes underlying
this percolation transition, one must follow the reaction kinetics,
compare results with experiments directly measuring rheological properties
in the course of the polymerization reaction, and also critically examine
the steady state fluorescence technique employed. The ``rheo-kinetics" of
bulk polimerization is an extremely active and growing field of research,
where different considerations based on free-volume, entanglement (or
reptation), gel-effect and vitrification compete with and complement each
other.~\cite{rheokinetics} We will interpret our results in the light of
recent developments in the theory of the glass
transition.~\cite{Grest,Colby,Glotzer1,Glotzer2}

In section 2, we outline our experimental method, in section 3 we discuss
the various mechanisms which come into play as the reaction proceeds. In
section 4, we analyze the results in terms of percolation theory and
extract the scaling exponents.  In section 5, we present a discussion of
the results.

\section{The Experiment}

Our experimental method, the {\it in situ} monitoring of free-radical 
polymerization by a steady state fluorescence technique, is the same as
the one we used to  determine the critical exponents for the
gel fraction ($\beta$) and the average cluster size ($\gamma$) in a
previous paper~\cite{Yilmaz}, where we also found percolation 
exponents.~\cite{Yilmaz3}

Fluorescence measurements yield direct, in situ information regarding the
connectivity and available free volume in polymer melts and
gels.~\cite{OP0} The use of fluorescent probes to study the different
environments in which the probe molecule finds itself, including
experiments on polymerization~\cite{OP1,Serrano}, chemical gel
formation~\cite{OP2,OP3,OP4}, swelling~\cite{OP5} and slow release of 
large
molecules~\cite{OP7}is an extremely well established method. Such
measurements have also been applied to the study of glass formation,
especially by the group of Ediger~\cite{Ediger1}.

\subsection{Experimental procedure}

Pyrene was used as the fluorescence probe to detect the free-radical
polymerization process of poly(methymethacrylate) (PMMA), a linear
polymer. The characteristic time for the direct relaxation by fluorescent
emission of the pyrene molecules is $\sim 10^{-9}$ seconds, a time scale
much smaller than those involved in reptation, and of about the same order as Rouse
dynamics~\cite{Doi,Binder,deGennes,Edwards,Fuchs, Bennemann}. Thus, 
extremely fast ``snapshots" of the effective network can be
obtained as the polymerization reaction proceeds, 
without mechanically disturbing it, as one would have to do with
conventional viscosity measurements.

Our experimental setup is the same as in ref.~\cite{Yilmaz}, except for
the absence of a crosslinker, so that only linear polymers are obtained.
The free radical polymerization of MMA was performed in the bulk in the
presence of 2,2'-azobisisobutyronitrile (AIBN) as an initiator, at
60$^\circ$, 70$^\circ$, 75$^\circ$, $85^\circ$ and $125^\circ$ C.  The
glass transition temperature of bulk PMMA is $105^\circ$C for the average
chain lengths attained here. The pyrene molecule was excited at 345 nm and
the variation in the fluorescence emission intensity, $I$, was monitored
with the time-drive mode of the spectrometer, with a frequency of up to 10
points per second, staying at the 395 nm peak of the pyrene emission
spectrum.

\begin{figure}
\leavevmode
\rotatebox{0}{\scalebox{.3}{\includegraphics{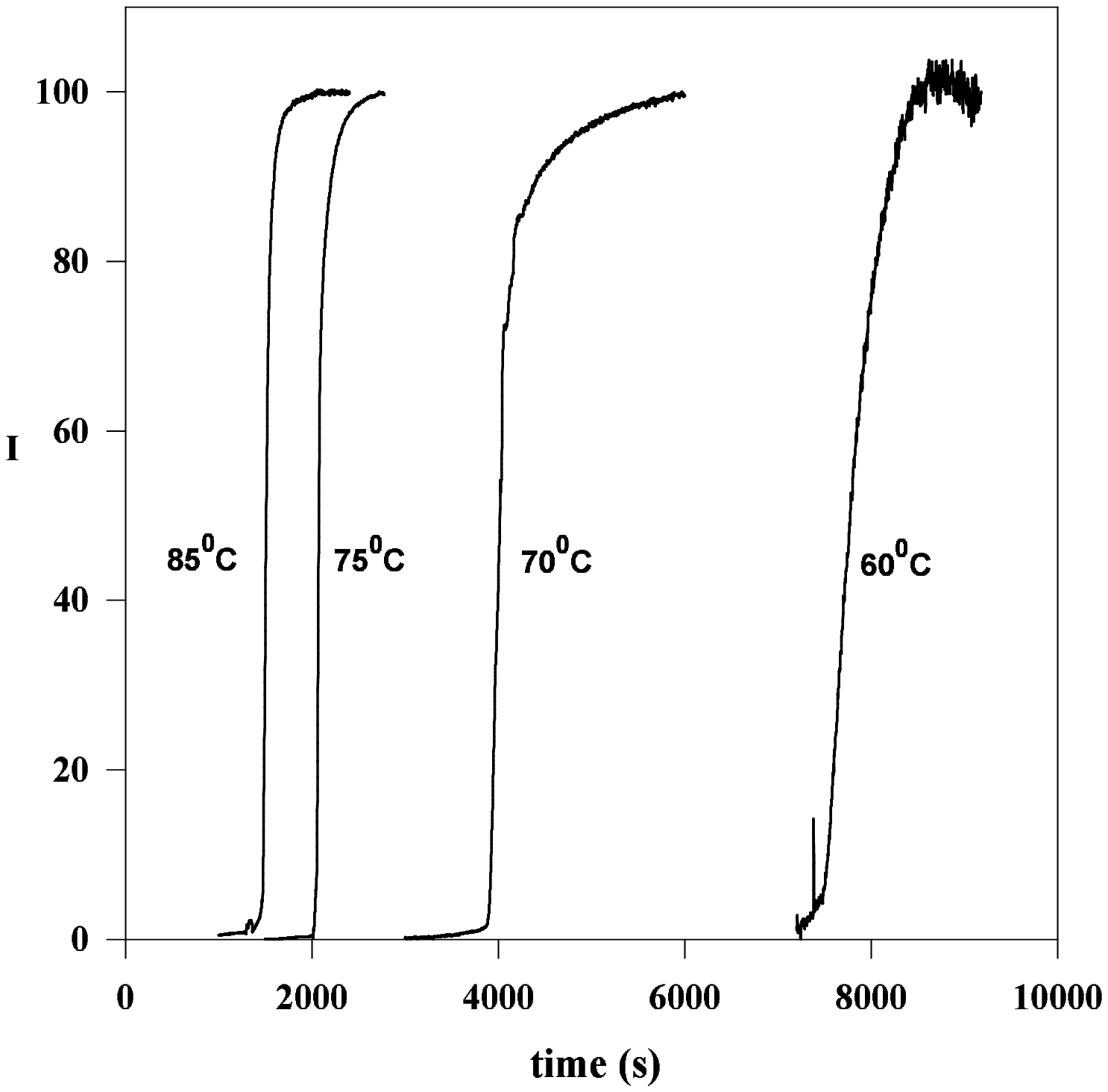}}}
\caption{The sharp increase 
in the fluorescence intensity observed as a function of time 
for different temperatures. The maximum intensity is normalized to 100.}
\end{figure}

\subsection{Results}
In Fig. 1, we present our results for the normalized pyrene fluorescence 
intensities as a function of the reaction times, for different temperatures. 
The results for $125^\circ$C are very similar; however due to 
the much greater mobility and hence higher reaction rates at the higher temperature, the transition occurs 
already at 300 seconds, and therefore is off-scale in this figure.

The end product is rubbery at $125^\circ$C and glassy otherwise.  In any
case, at the end of the polymerization reaction the bulk polymer does not
flow, but retains its shape.  The addition of solvent causes the polymer
to swell, as a gel would, up to volume ratios of around $1:1$.  This 
 indicates that the sample has definitely undergone a transition to a gel 
phase, due to the
physical entanglements of the linear chains. Larger quantities of solvent
dissolves the top layers, until, at a polymer to solvent ratio of about
$1: 2.5\pm 0.2$ the whole sample becomes liquid. Since we started off with
a bulk polymer, this ratio corresponds to the volume ratio at which the
chains disentangle. This is in the same ballpark although a bit bigger
than the conversion estimated from the value of $I$ at the inflection
point in Fig. 1, from $p_e\equiv I_e/I_{\rm max}\simeq 0.3$
(see~\cite{Yilmaz2})and the percolation threshold for a cubic lattice in
three dimensions $(p_c=0.31)$~\cite{Stauffer1}. For a cubic lattice,
Kantor and Hassold~\cite{Kantor} find an ``entanglement threshold" very
close to the ordinary percolation threshold, $p_e\simeq p_c-2\times
10^{-7}$. (Also see~\cite{Holroyd1,Holroyd2,Grimmet}. It should be noted,
however, that these authors consider clusters including branch points as
well as those that cannot be pulled apart because of topological
constraints.)

These considerations seem to point in the direction of a gelation
(entanglement) transition~\cite{Fetters,Kholodenko,Binder,Paul1,Paul2} 
at the onset 
of the sharp rise, ending, however in a glassy phase, within a
very small portion (duration wise) of the whole polymerization process.
The glass transition temperature is swept up as the polymerization
proceeds, until it crosses the temperature of the sample.~\cite{TmMMA}
Even in the case of the run with the temperature held at 125$^\circ$C, one
is within about 20 degrees of the glass transition temperature, and one
has to take into account the precursor effects that may be present.  
Qualitatively very similar behaviour observed many tens of degrees 
above the glass transition temperature~\cite{Serrano}, remains, 
however, a puzzle.

\section{Reaction kinetics and rheology}

A close examination of the reaction kinetics of the free-radical
polymerization process~\cite{Okay1,Okay2,Okay3,Zhu} reveals that in the
initial stages (the flat, vanishingly small parts in Fig.1, preceding the
rapid rise) the release of the free-radical which initiates chain
formation, and the self- or cross-termination of the chains, balance each
other out. Thus the chains which are formed are all more or less of the
same length; i.e., as conversion proceeds at a fairly constant rate, the
lengths of the chains that are formed are not growing on the average. As
more of the monomers are converted, we reach a point where the
fluorescence intensity starts picking up.  This coincides with the rise in
viscosity of the medium, with a concomitant rise in the rate of reaction,
with the trapping of the free radicals\cite{Okay3,Zhu,Torkelson}, termed 
the ``gel
effect." (Some authors refer only to the point where the reaction rate
reaches a maximum, as the ``gel effect"~\cite{Okay3}). Somewhat
counterintuitively, the reaction rate grows as the reaction becomes
diffusion limited, because end to end- or self-termination of the chains
become increasingly rare and the length of the growing chains can become
very large as a result. We will argue in this section that the fluorescent
signal grows because of the trapping of the pyrene probes as well as the
free radicals, in regions of relatively high density of the polymeric
chains, where the cage effect comes into play.~\cite{Zhu,Tian,Ediger1}

The curves in Fig. 1 are quite typical of polymerization reactions with or
without ~\cite{OP1,Serrano,Yilmaz,Yilmaz3,Okay3,Zhu,Tian} crosslinkers,
and display a very sharp rise, with an inflection point, where the time
derivative of the curve passes through a maximum, as shown in Fig. 2. (The
duration and slope of the initial region of low conversion may depend on
the temperature and weight fraction of the crosslinker or initiator.)

In an experiment involving a free-radical co-polymerization with cross
linkers, Okay et al~\cite{Okay3} have monitored the viscosity of the
medium using dilatometric techniques, and found that the gel point is
located roughly at the onset of the gradual rise in the pyrene intensity
curve, and definitely before the inflection point is reached. In the
present experiment, since no crosslinkers are used, the increase in the
viscosity may be ascribed to the physical entanglement of the polymer
chains~\cite{Zhu}, giving rise to a purely geometrical gel phase (note
that the unconverted monomers are easy to displace and provide the ``sol''
phase). The PMMA chains that are formed interact very weakly with each
other, so that inter-chain attraction as in the formation of
thermoreversible physical gels~\cite{Saiani,Erukhimovich} is less
important here.

While chemical gelation in the presence of chemical cross-linkers exhibits a 
sharp transition which 
is modelled by percolation 
theory~\cite{Stauffer1,Stauffer2,Herrmann1,Herrmann2}, the formation of physical 
gels by bulk linear polymers is less well understood. It may be conjectured on the basis of scaling arguments and renormalization group calculations~\cite{Kantor,Duygu} that this gel point is also controlled by percolation exponents, although it arises purely from entanglement effects and there are no (chemically bonded) branch points in the spanning network thus formed. Nevertheless, the above dilatometric findings on a similar system (albeit having crosslinkers) suggest that the pyrene fluorescence is probing the post-gel region and not the vicinity of the gel point.  Further evidence for this is provided by a critical examination of the mechanism for the unquenching of the fluorescence itself.

The steady state fluorescence technique relies on the fact that the
relaxation of excited aromatic molecules via indirect non-radiative
transfer of energy to the high--frequency vibrational modes of the small
solvent particles or monomers is forbidden~\cite{Jones} with the increased
viscosity of the medium. The rates of intra-molecular or inter-molecular
non-radiative transitions are determined by the availability of
appropriately spaced levels either in the molecule itself or in the
surrounding molecules. These levels are altered with the degree of
conversion of the polymer melt, due to the local dynamical constraints
arising from a reduction in the available free volume, and the consequent
slowing down of segmental motions, leading to
vitrification.~\cite{Ediger1,Ediger2,Ediger3,Ediger4, Ediger5,Ediger6}
In this context, free volume will mean the volume occupied by unconverted 
MMA molecules.

The connectivity may also contribute indirectly to retarding the local
dynamics, in the present case through entanglement effects (since no
chemical cross linkers are present). But the crossover between Rouse to
reptation dynamics takes place above the glass transition (in temperature,
or free volume).  Moreover, it is generally believed that the cage
effect~\cite{Goetze} leading to this slowing down and vitrification occurs
at much smaller spatial scales (in fact of the order of the polymeric
bonds)~\cite{Fredrickson} than those of Rouse or
reptation~\cite{Fuchs,Bennemann} dynamics.

\begin{figure}
\leavevmode
\rotatebox{0}{\scalebox{.3}{\includegraphics{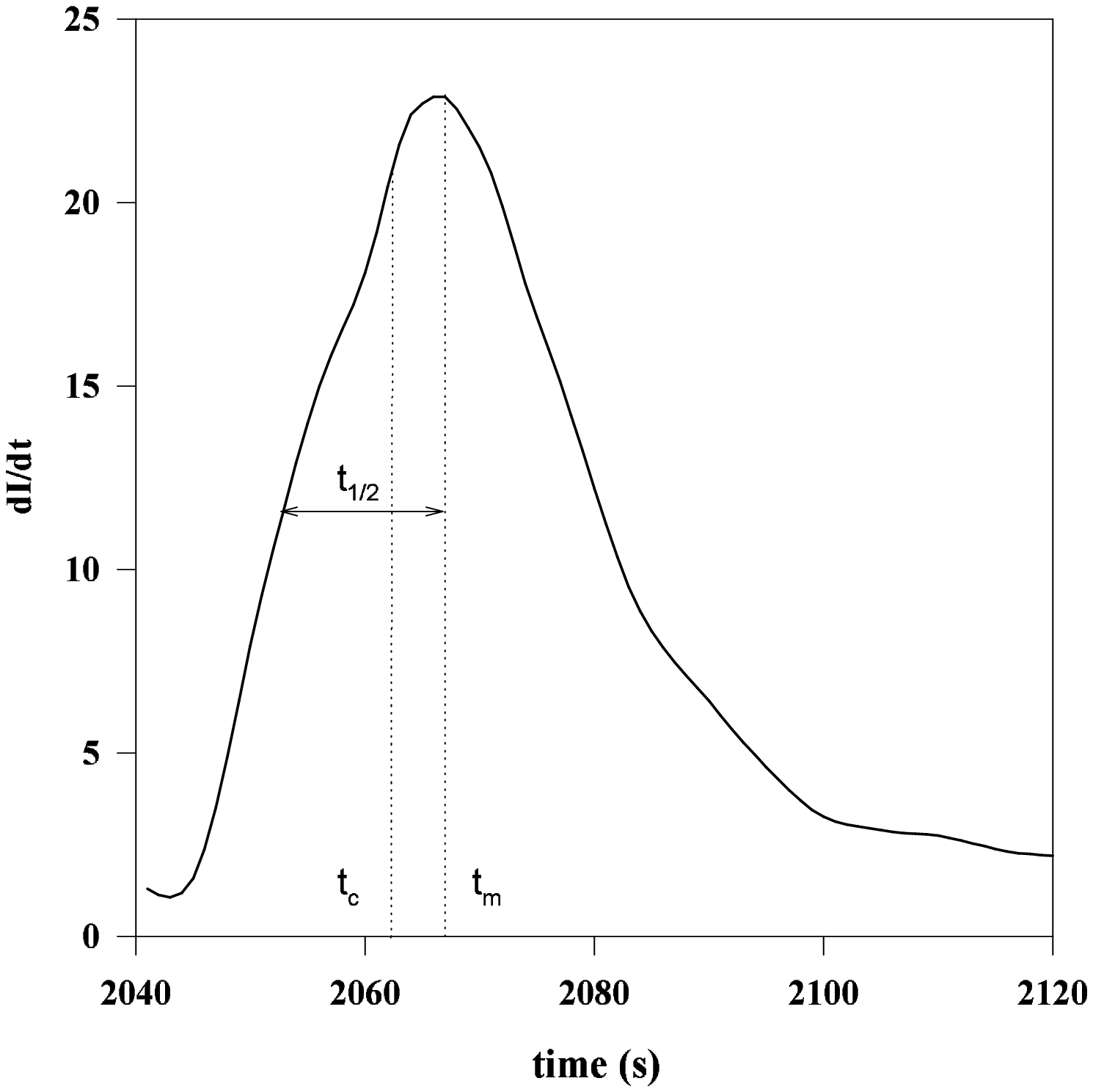}}}
\caption{The typical critical peak exhibited by the
 derivative of the fluorescence curve, at $75^\circ$C, in the vicinity of the 
critical point. 
The maximum is at $t_m$, the half width at half maximum, $t_{1/2}$, and
the critical point, $t_c$ are indicated.}
\end{figure}

We can make a rough estimate of $M_c$, the average length of the chains at
the inflection point, from the value of the conversion fraction
$p_e \simeq 0.3$~\cite{Yilmaz2}, assuming
that the number of chains is approximately the same as the number of
initiators.  The molecular weights of the MMA and AIBN molecules are 100
and 164, respectively, and the experiments were performed with samples
consisting of 1ml MMA (density 0.96 gr/ml) and 0.0026 gr AIBN. We find
$M_c\sim 180$, which is of the same order of magnitude as the chain length
at which the chain dynamics crosses over from Rouse-like to reptation
behaviour~\cite{deGennes,Doi,Edwards}, as estimated by Kreer et
al.~\cite{Binder} from simulations. This is an order of magnitude larger
than an alternative estimate~\cite{Paul1,Paul2} which can be obtained by
comparing the average end to end distance to the diameter of the reptation
``tube." 

On the other hand, the time scales involved in the cage effect are of the order of pico seconds, while in Rouse or reptation motion they range from nano to microseconds~\cite{Fuchs}. Therefore it would seem that those degrees of freedom which would be frozen out with respect to the pyrene probe with a time scale of the order of nanoseconds, would be the latter.
Thus it is difficult to unequivocally identify the contributions to the 
unquenching of the pyrene fluorescence, as arising strictly from the connectivity (entanglements) or from the formation of density inhomogeneities that are the precursors of glass formation, merely from  considerations of time and length scales. 

Further information is provided by the ESR spectra in the course of the
polymerization (free radical copolymerization of MMA and ethylene glycol
dimethylacrylate - EGDMA)~\cite{Tian}.  These measurements reveal that the
radical concentration exhibits a sharp peak right at the inflection point
of the polymerizations curves, which are exactly similar to those
appearing in Fig.1. Even more crucial is the fact that, in the region
between the inflection point and where the curves flatten off, the ESR
spectra switch from having 13 lines to 9 lines; with the 9 line spectra
being contributed by radicals finding themselves in solid, i.e., glassy
environments.~\cite{Zhu,Tian,Klooster1,Klooster2} A
simulation~\cite{Klooster1} based on a random walk percolation
model~\cite{Herrmann1,Herrmann2,Pandey} compares resonably well with the
experiment below the percolation point, but fails to do so beyond the gel
point, occuring at about weight 30\% polymerization.~\cite{Klooster1}

These considerations lead us to conjecture that the critical scaling
behaviour which we observe in the vicinity of the inflection point of
Fig.1, and which is discussed in the next section, arises from the
percolation of dense regions where the cage effect can be observed, and
vitrification sets in.
This also raises the possibility in Ref. ~\cite{Yilmaz}, the critical
scaling that is observed is due to the percolation of slow regions in this
sense and may lead us to reasses our interpretation of the latter
results.

\section{Percolation of slow regions}

We would like to cast the quasi--static properties of the polymer melt in
the language of percolation, to interpret our results.

Note that free monomers are relatively easy-to-displace 
objects which do not constrain the
motion of the large chains. 
By a slow region, we shall mean a region of
sufficiently high density, such that parts of several chains, or even of
the same chain, get in each other's way and inhibit each other's motion on
a short time scale. 

We take the volume fraction occupied by the total
number of monomers incoporated into the chains as the 
``occupation probability,'' $p$, of the sites of a three dimensional
lattice.  We will take a cluster to be  a set of occupied points
on the lattice, which are nearest neighbor to at least one other member of
the set.~\cite{Stauffer1}  Therefore such nearest neighbor occupied sites may be
considered as belonging to the same cluster, {\em regardles of whether they are
chemically connected} (i.e., belong to the same chain) or not. A spanning
cluster of this description will be called a percolating cluster. We then argue that the pyrene molecules contributing to $I$, the
fluorescent intensity, are precisely those which are trapped within
interstitial regions of these clusters.

If $M$ is the average molecular weight of the chains, 
\be p= MCv_0/V\equiv v_0/v\;\;\;,\label{p}\ee 
where $C$ is the total number of chains,
$v_0=V/N$ is the volume occupied by a single monomer, and $N$ is the
total number of monomers. The average length $M$ of the
chains obeys the differential equation
\be
{d M(t) \over dt} = k(N-CM)\;\;\;\;,\label{diff}\ee
over a sufficiently short time interval where 
the time dependence~\cite{Yilmaz2} of the reaction rate $k$ due to the change in 
the viscosity, and of $C$ (due to the gradual dissolution and termination of the 
initiators) can be neglected. 
Thus, for  relatively short time intervals, we get a linear growth law, 
\be M(t)-M(t_0)= [M_\infty-M(t_0)]\; kC(t-t_0)\;\;\;.\label{M(t)}\ee
where  $M_\infty=N/C$ and $t_0$ is an arbitrary starting point. 
Defining $t_c$ as the time at which the entanglement percolation 
threshold $p_e$ is reached, and using Eq.(\ref{p}), we have,  
for sufficiently small  $\vert t-t_c\vert $, 
\be \vert t-t_c\vert ={1\over  kC}{\vert p-p_e \vert \over 1-
p_e}\propto \vert p-p_e \vert\;\;\;.\label{A1}\ee

Since below $t_c$, there is no percolating cluster, the total 
normalized
fluorescent intensity will be proportional to the average cluster
size $S$. For $t>t_c$, most of the
pyrene molecules are trapped in the macroscopic network of slow regions, 
and $I$
then measures the fraction $P_\infty$ of the monomers
that belong to the macroscopic  cluster.~\cite{Yilmaz}
For a system with linear size $L$, 
the scaling forms for the quantities $S$ and $P_\infty$ around the
percolation threshold, together with (\ref{A1}) yields,
\begin{eqnarray}
I\propto \cases{  S \sim  (t_c-t)^{-\gamma},  & $t<t_c $ \cr
L^{d_f} P_{\infty} \sim L^{d_f} (t-t_c)^\beta, & $t>t_c\;\;.$ }
\label{scale} \end{eqnarray}
Here, $\beta$  and $\gamma$ are the  critical exponents for
the strength of the infinite cluster and the average cluster size,
 and $d_f=(\beta +\gamma)/\nu$
is the fractal dimension of the spanning cluster, with 
$\nu$ being the correlation length exponent.~\cite{Stauffer1}
Notice that we need not subtract the value of
$I(t_c)$  from $I(t)$ in (\ref{scale}) for $t>t_c$ since we are assuming
that once the threshold has been crossed, the unquenched fluorescence
intensity is being contributed essentially by the monomers
trapped in the incipient infinite cluster.

The time derivative of the intensity $I$ is plotted in Fig. 2, and looks like
a typical critical peak, with rounding due to ``finite size" effects.   
The fits to the double logarithmic 
plots of the
fluorescence intensity v.s. $\vert t-t_c\vert $ for $t>t_c$ and
$t<t_c$, are shown in Fig. 3.
The critical point $t_c$ 
and the exponents $\beta$ and $\gamma$
are determined  by simultaneously fitting 
the intensity data  to the behaviour 
in Eq.(\ref{scale}), 
in such a way as to obtain the greatest range in $\vert t-t_c\vert$ 
over which one finds scaling, 
both above and below the critical point.~\cite{Yilmaz}
In this way it is possible to remove the ambiguity from the position of the 
critical point and greatly enchance the accuracy of the critical exponents.
We see that the critical region is confined to 
the interval $3<\vert t-t_c \vert < 30$.
The values of the exponents are given in Table I.  
The agreement with the known values ($\beta = 0.41,
\gamma=1.80$) of the
percolation exponents in three dimensions, is very good,
with
$\beta=0.41\pm 0.01$ and $\gamma =1.7\pm 0.1$. The error bars are
estimated from the values reported in the table.  The
numbers for $60^\circ$C differ markedly from the rest and 
we have excluded 
them  from our estimate of the exponents. 
At such low temperatures 
the mobility of the MMA are reduced~\cite{Serrano} to 
the point where the reaction ends before 
all the monomers are exhausted (the absolute intensity is lower 
at this temperature). The remaining monomers act as if a solvent were 
added to the bulk, increasing the lower length scale of the 
percolating network~\cite{Yilmaz}, and distorting the scaling behaviour.
On the other hand, at 125$^\circ$C, the sample is boiling before 
the polymerisation has had time to progress, and  the curve 
below the transition point  is rather  noisy, so that  we 
find $\gamma= 1.4$; we recover  $\beta=0.41$ above the transition point.

\begin{table}
\begin{ruledtabular}
\caption{Experimentally determined values of $\beta$ and $\gamma$.
}
\begin{tabular}{c|c|c|c|c|c|c} 
{\boldmath ~$T(^\circ C)$~}&60 &    70  & 75     & 80    & 85   & 125 \\ 
\hline\hline
$\gamma$       &0.40&   1.8  &  1.6   &  --   & 1.71  & 1.41 \\
$\beta$        &0.45&   0.42 &   0.41 &  0.41 &  0.41  & 0.41\\
\end{tabular}
\end{ruledtabular}
\end{table}

The critical point $t_c$ obtained from the best 
scaling fits is smaller than $t_m$ marking the maximum 
of the $dI/dt$ curve.  
The quantity $c\equiv (t_m-t_c)/t_{1/2}$ where $t_{1/2}$ is the half--width
of the
$dI/dt$ curve at half maximum, is found to be 
\be
c\equiv {t_m-t_c \over t_{1/2}} = 0.248\pm 0.005   \;\;,
\label{formula}\ee
over the whole range of experiments, and for different temperatures.  
This value is the same as that found from the measurements
performed near the percolation transition found from fluorescence measurements on crosslinked
gels~\cite{Yilmaz}.  

Finite size scaling~\cite{Stauffer1} predicts
that this ratio 
is in fact a universal constant for percolation.
From Eq.(\ref{scale}) we see that 
$dI/dt$ must obey  the finite size scaling relations
\be
{dI \over dt}\sim L^{(\gamma +1)/\nu}\cases{
\phi_s^\prime[(t_c-t)L^{1/\nu}]
& $t< t_c$ \cr
 \phi_p^\prime[(t-t_c) L^{1/\nu}] & $t>t_c$}\ee
with the scaling functions $\phi_s(x)\sim x^{-\gamma}$ and $\phi_p(x)\sim
x^{\beta}$ for $x\gg 1$, and $\sim$ const. for $x\ll 1$. Let us define the
universal scaling function $\Phi^\prime$ by $dI/dt \sim L^{(\gamma
+1)/\nu} \Phi^\prime[(t-t_c) L^{1/\nu}]\;\;.$ Then $\Phi^\prime$ clearly
corresponds to the curve depicted in Fig.2, both above and below $t_c$,
and $c$ must be given in terms of its moments, $z_n = \int z^n
\Phi^\prime(z) dz$, as $c=z_1/(z_2-z_1^2)^{1/2}\sim O(1)$. The fact that
we find the same numerical value for $c$ as that determined from the
cross-linking gelation experiment~\cite{Yilmaz} is further indication that
the  transition we observe here is in the same universality
class as percolation.

\begin{figure}
\leavevmode
\rotatebox{0}{\scalebox{.3}{\includegraphics{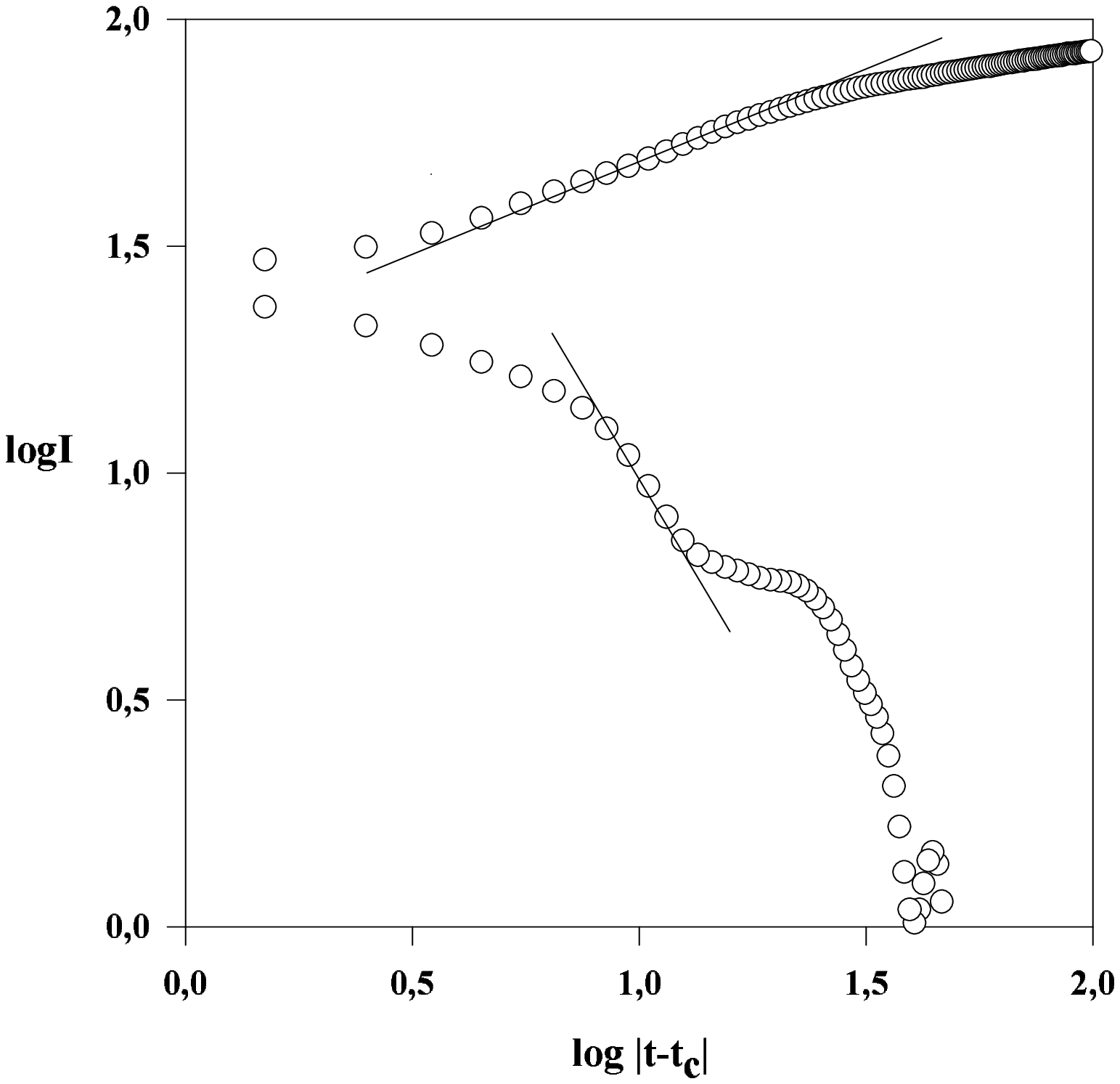}}}
\caption{Best double logarithmic fits to the curve for $75^\circ$C shown in
Fig. 2, yielding $\gamma = 1.60$ (lower curve) and 
$\beta=0.41$ (upper curve). Every tenth data point is plotted for greater 
clarity.}
\end{figure}      

To understand this universality, we note that a dense region will
look like a branching point under a coarse graining by some factor $b$,
(while a bigger ``blob" will now look like a tight entanglement region).  
If the percolating entanglement cluster is self-similar at length scales
$\ell \ll \xi$, where $\xi$ is the correlation length diverging like
$\vert p-p_e\vert ^{-\nu}$, near the  percolation threshold,
this transformation will simply renormalize $\vert p-p_e\vert $ by a
factor $b^{1/\nu}$, but otherwise leave the scaling behaviour unaffected.
Below the percolation threshold, the finite clusters themselves are
fractal for $\ell \ll \xi$ and the same argument applies.

\section{Discussion}

We have observed that steady state fluorescence experiments in a polymeric
melt reveal a sharp transition in the narrow crossover region from low
conversion to high conversion and vitrification. The transition is
characterised by percolation exponents.  On the basis of arguments
presented in section 3, we believe that the critical scaling behaviour we
observe in the fluorescent signal as a function of time is not due to the
physical gelation of the sample due to entanglements, but to the
percolation of dense regions where the cage effect constrains the local
segmental motions of the linear chains and the radicals as well as the
pyrene probe.

It is relevant to note in this context the dynamical scaling
hypothesis~\cite{Colby}, which asserts that, as the glass transition is
approached from above, long range dynamical spatial correlations build up
between particles of fragile glass formers. Dynamical clusters are formed
out of those objects which have cooperatively rearranged, using the same
free volume to accomplish the move.~\cite{Colby,Glotzer1,Glotzer2} In the
present context, this idea acquires particular significance, because with
the onset of the ``gel effect,'' and the trapping of the free radicals,
further polymerization must proceed in exactly the same way as the motion
of a particle in a vitrifying medium, leaving behind, moreover, a
permanent ``trail" consisting of a long chain that is formed as a result.  
The chain can only grow if unattached monomers (which act like free volume
with respect to monomers already incorporated into the chains) can
exchange places with segments of neighboring chains trapping the radical
at the head of the growing chain. Once a monomer is attached to the
growing chain, a new monomer has now to exchange places with the
surrounding chain segments, and will in turn be eaten up by the radical,
etc. This process then leaves behind a growing chain confined within dense
regions, and interconnecting these dense regions.  The sudden rise in the
rate of polymerization at the onset of the ``gel effect" feeds on itself
in nonlinear fashion, and suddenly gives rise to the percolation of dense,
slow regions in the melt.  The entrapment of the pyrene probe in these
slow regions gives rise to the critical behaviour of the fluoresent
intensity.

{\bf Acknowledgements}

It is our great pleasure to acknowledge a very useful discussion with
Prof. O. Okay. Y.Y., A.E. and \"O. P. would like to thank the Turkish 
Academy of
Sciences for partial support.

\end{document}